\documentclass[aps,pra,10pt,twocolumn]{revtex4-2}
\usepackage{graphicx}
\usepackage{amsmath,amsthm,amssymb,dsfont}
\usepackage{mdframed}

\usepackage{subcaption}
\usepackage{dcolumn}% Align table columns on decimal point
\usepackage{bm}% bold math
\usepackage{braket}
\usepackage{comment}
\usepackage{float}
\usepackage{url}
\usepackage{ulem}

\begin{document}

\preprint{}

%%--
\title{Characterization of non-classical particle propagation \\using superpositions of position and momentum}

\author{Yuki Senoo}
\author{Holger F. Hofmann}
\author{Hiroki Yamakami}
\author{Masataka Iinuma}
\affiliation{%
 Graduate School of Advanced Science and Engineering,
 Hiroshima University,\\Kagamiyama 1-3-1, Higashi Hiroshima, 739-8530, Japan
}

%%--
\begin{abstract}
The uncertainty principle suggests a quantitative trade-off between the control of position and the control of momentum in particle propagation. However, a superposition of two states with very different uncertainty trade-offs introduces an interference term that seems to combine precise statements about position and about momentum, allowing us to study how quantum mechanics describes the propagation of individual particles in free space. Here, we present a detailed experimental study of photons prepared in a superposition of position and momentum generated in a Sagnac interferometer. The transverse distribution of photons was obtained with three different measurement settings at the output port of the interferometer, corresponding to the initial position distribution, the initial momentum distribution, and an intermediate propagation time at which the contributions of initial position and momentum uncertainties are approximately equal to each other. We show that the interference effect localizes the photons in narrow intervals of position and momentum, resulting in a quantitative violation of Newton's first law as the interference pattern spreads out at the intermediate position. 
The data obtained can be used to demonstrate the negativity of the Wigner function in regions outside the position and momentum intervals in which the position and momentum contributions are confined.
\end{abstract}

\maketitle

\section{\label{sec:level1}Introduction}

One of the most fundamental problems of quantum mechanics is the relation between wave propagation and the detection of individual particles. In general, quantum interference makes it impossible to trace back the path taken by a particle through an interferometer \cite{wootters,scully,danan,Zhou,Lev,OptQuant}. It might seem that the problem does not appear in free space, since most beam shapes seem to be consistent with propagation in a straight line, and this is borne out in the positivity of the Wigner function of such states, which can be interpreted as a joint probability of position and momentum \cite{wigner,moyal}. However, Wigner functions can also be negative, proving that quantum interference modifies the motion of a particle in a fundamental manner. 

As the example of the Wigner function shows, the main problem is the relation between the statistics of position and momentum at one time and the position distribution at a different time. Early on, it was noted that this problem seems to be similar to the problem of particle statistics in hydrodynamics, where the Schr\"odinger equation can be interpreted as a complex diffusion equation \cite{madelung}. Later, Bohmian mechanics explained the difference by introducing a quantum potential defined by the wavefunction, effectively replacing free space motion with dynamics modified by additional forces \cite{bohm,scherer}. Interestingly, the velocities associated with Bohmian mechanics can be identified with weak values of momentum \cite{aharonov,wiseman,Kocsis,blikoh}. However, this approach abandons the phase space symmetry of position and momentum that characterizes the Wigner function \cite{schleich}. The experimental evidence provided by weak values should therefore be treated with caution. In fact, a direct investigation of the joint statistics of position and momentum using weak measurements results in an experimental characterization of the Kirkwood distribution, an alternative to the Wigner function that is more closely related to the wavefunction of the particle \cite{kirkwood,lund}. The quantum formalism itself seems to suggest that quantum interference effects are a fundamental part of particle propagation that cannot be reconciled with the assumption of specific trajectories. Several experimental investigations of particle trajectories have therefore focused on interferences between different paths \cite{shinha,sawant,omar}. However, these experiments do not characterize the trajectories of individual particles. 

The difficulty of identifying particle trajectories in quantum mechanics is best illustrated by the negativity of the Wigner function. As pointed out by Moyal \cite{moyal}, the Wigner function is the only quasi probability of position and momentum that is consistent with straight line motion in free space. Negative Wigner functions therefore represent violations of Newton's first law, indicating that classical concepts of motion are not valid in quantum mechanics. Instead of introducing modified laws of motion, it may therefore be better to investigate the specific violations of classical expectations associated with the negative values of Wigner functions. One phenomenon that demonstrates such a violation is the quantum backflow \cite{bracken,yearsley,backflow}, where it is shown that a probability current flows in a direction opposite to that of the available values of momentum. Specifically, it was shown that a superposition of momentum eigenstates with $p>0$ can describe a probability current from $x>0$ to $x<0$, even though this would seem to require negative momentum. This effect clearly demonstrates that the classical constraints that a momentum distribution would impose on the time evolution of the spatial distribution are not valid. 

A closely related quantum phenomenon is the violation of an inequality by a superposition of position and momentum \cite{hofmann,hofmann3}. This inequality relates the probability of finding a particle in the position interval $L$ and the probability of finding a particle in the momentum interval $B$ to the probability of finding the particle in a position interval $M$ at a later time, where the classical laws of motion would require that all particles in both $L$ and $B$ necessarily arrive at $M$. The corresponding inequality relating the probabilities $P(L)$ and $P(B)$ to $P(M)$ can be violated by a superposition of a state localized in $L$ and a state localized in $B$, demonstrating that particles in this superposition state cannot be traveling along straight lines. The predicted violation has been confirmed experimentally in a setup generating an equal superposition of position and momentum \cite{takafumi}. Here, we present new experimental data obtained in a similar setup in order to study the mechanism by which quantum interference modifies the propagation of particles. 

In the following, we obtain detailed count rates for the transverse spatial distribution of photons in position, in momentum, and at the intermediate position at which the maximal inequality violation is observed. We identify the contributions from the position state, from the momentum state, and from the interference between the two. We find that the state can be reconstructed from the experimental data, with an effective quasi probability distribution of $w_L=0.355$ in the position state, $w_B=0.493$ in the momentum state, and $w_{\mathrm{inter}}=0.152$ in the interference term. The interference term is responsible for an overall violation of the inequality by $0.060$, with positive contributions of $15.2 \%$ to both $P(L)$ and $P(B)$, and an interference pattern that contributes only $5.4 \%$ to $P(M)$. We also show that this inequality violation corresponds to negative values of the Wigner function in the regions outside of the intervals $L$ and $B$, where the inequality violation corresponds to the total negativity obtained from integrals over the phase space regions outside of the intervals. This negativity of the Wigner function is a direct consequence of the simultaneous control of position and momentum expressed by the contributions of the interference term to $P(L)$ and $P(B)$. Our results show how interference effects reshape the propagation of quantum particles in free space, providing important quantitative evidence for the failure of the classical particle picture in quantum mechanics.

The rest of the paper is organized as follows. In Sec. \ref{sec:level2}, we introduce the concept of experiment and the explanation of the inequality. The relation between the inequality violation and the negativity of the Wiger function is also included. The experimental setup and the procedure are explained in Sec. \ref{sec:level3}, and the measurement results and the numerical analysis are discussed in Sec. \ref{sec:level4}. We evaluate and isolate the quantum interference appearing in spatial distributions based on quasi probability, and discuss the comparison between the initial and intermediate times. This analysis enable us to directly discuss to the contribution of the interference term to the negativity of Wigner function. Finally, we show our conclusion in Sec. \ref{sec:level5}.

\section{\label{sec:level2}Theoretical background}
\subsection{The particle propagation paradox}
\begin{figure}[h]
    \centering
    \includegraphics[width=0.9\linewidth]{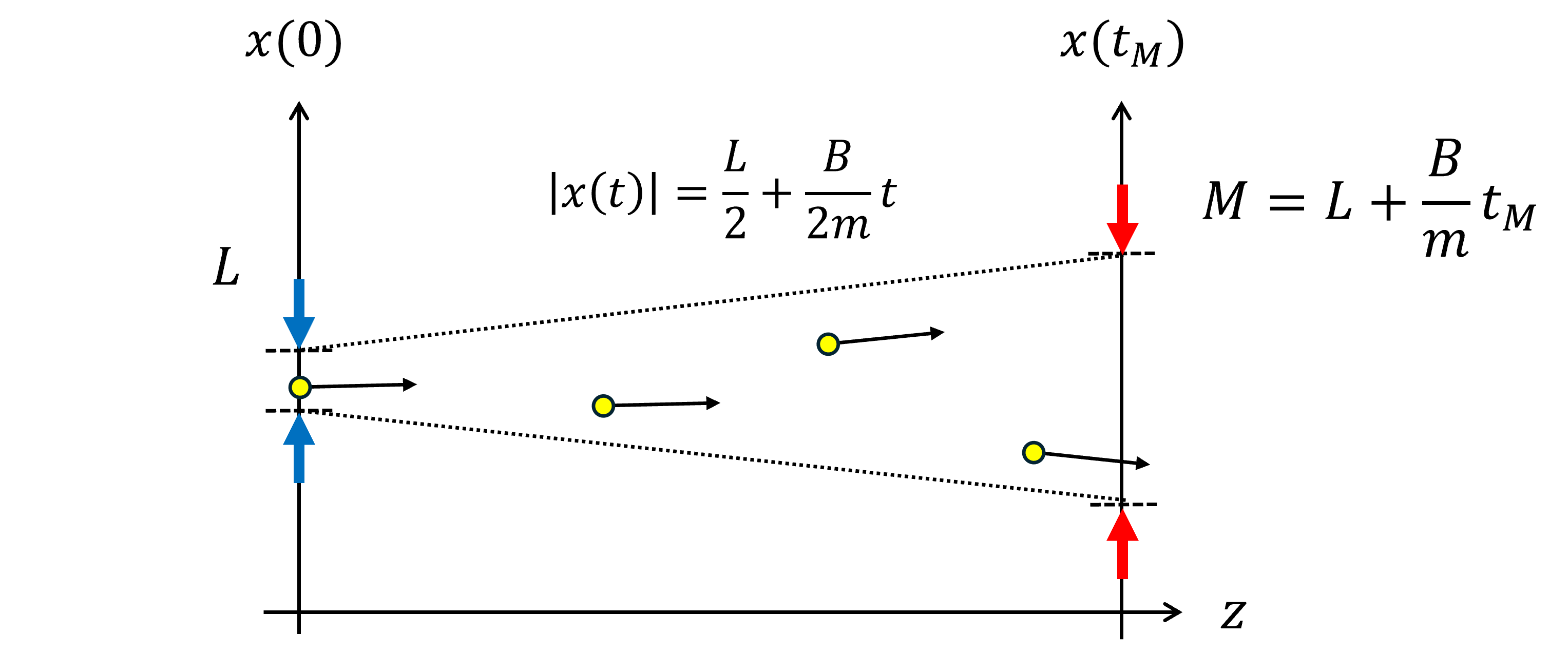}
    \captionsetup{justification=raggedright,singlelinecheck=false}
    \caption{
    Experimental test of motion in free space. If particles travel in a straight line, any particle that can be found simultaneously within a position interval $L$ and a momentum interval $B$ at $t=0$ will not have traveled beyond a classically expected bound of $L+(B\:t_M)/m$ at time $t_M$.
    }
    \label{fig:concept}
\end{figure}

We consider the motion of massless particles (photons) propagating in free space as shown in Fig. \ref{fig:concept}.  
The velocity along the z-direction is close to the speed of light, so that the distance along the z-axis can be identified with the time of propagation, $t=z/c$. The time evolution of the transverse position $x(t)$ is given by the transverse velocity $v=\theta c$, where $\theta \ll 1$ is the angle at which the photons propagate relative to the z-axis. The momentum corresponding to this angle is given by $\theta$ times the total momentum of the photon and the ratio of velocity and momentum is given by the mass of the propagating photon, $m=h/(c\lambda)$. The motion in the transverse $x$ direction thus corresponds to the free space motion of a particle of mass $m$, where the passage of time is recorded by the motion along the z-axis.

Newton's first law states that particles in free space will not change their velocity. This means that particles of mass $m$ with positions in an interval of $L$ around $x=0$ and a momentum interval $B$ around $p=0$ at $t=0$ should pass through positions in the interval $|x\left(t_M \right)| \leq M/2$ at time $t=t_M$, where $M$ is given by $L+(B/m) t_M$. It is therefore possible to test Newton's first law by comparing only three probabilities observed at three different times. These probabilities are defined as follows. 
$P(L)$ is the probability of finding particles at $t = 0$ in the position interval of width $L$ around $x=0$. Similarly, $P(B)$ is the probability of finding particles at $t = 0$ in the momentum interval of width $B$ around $p=0$. Since $p/m$ is the transverse component of the velocity of the particle, the momentum value corresponds to the angle $\theta=p/(mc)$ at which the photons propagate relative to the z-axis. In principle, this angle can be observed in space as soon as sufficient time has passed so that the transverse position is approximately given by $\theta ct$. Ideally, this condition is satisfied for $t \rightarrow \infty$, where the angles $\theta$ are mapped to macroscopic length scales defined by $ct$. In the following, a lens will be used to map the different angles $\theta$ onto different transverse positions, so that the ideal condition of $t \rightarrow \infty$ can be observed at a finite spatial scale. We can then identify the transverse momentum $p$ at $t=0$ from the position detected in the focal plane of the lens. Finally, the probability $P(M)$ is the probability of finding the particle in the position interval $M$ around $x(t_M)=0$ at time $t=t_M$, where $M=L+Bt_M/m$. 

Since any particle with a position in $L$ and a momentum in $B$ at $t=0$ will arrive within the interval $M$ at time $t_M$, the probability $P(M)$ has a lower bound of 
\begin{equation} 
P\left(M \right) \geq P\left(L\: \mathrm{AND} \:B \right), 
\label{eqn:newton_law}
\end{equation}
where the $P\left(L\: \mathrm{AND} \:B \right)$ is the probability of particles satisfying both $|x|\leq L/2$ and $|p|\leq B/2$. If $P(L)$ and $P(B)$ are known separately, the minimal value of $P\left(L\: \mathrm{AND} \:B \right)$ is given by $P(L)+P(B)-1$, obtained under the condition that all particles must either be in $L$ or in $B$. 
%%%
%%%
We can then formulate the inequality \cite{hofmann}
\begin{equation}
P(M)\geq P(L)+P(B)-1
\end{equation}
Any violation of this inequality proves that quantum particles do not move in a straight line. The minimal number of particles that do not obey Newton's first law is given by the defect probability,
\begin{equation}
P_{\mathrm{defect}} = P\left(L \right)+P\left(B \right)-1-P\left(M \right)
\label{eqn:P_defect}
\end{equation}
It has been shown that this inequality can be violated by particles in a superposition state of $\Ket{L}$ and $\Ket{B}$. Here, $\Ket{L}$ is a state representing the particles are in the position interval $L$ around $x = 0$ and $\Ket{B}$ is a state representing the particles are in the momentum interval $B$ around $p = 0$. The wavefunctions in position and momentum can be given as
\begin{align*}
\Braket{x|L} = 
\begin{cases}
\frac{1}{\sqrt{L}} & \mbox{for} \;\; |x|\leq L/2\\
\;\; 0 & \mathrm{else}
\end{cases}
\end{align*}
\begin{align}
\label{eq:wave}
\Braket{p|B} = 
\begin{cases}
\frac{1}{\sqrt{B}} & \mbox{for} \;\; |p|\leq B/2\\
\;\;0 & \mathrm{else}
\end{cases}
\end{align}
Since $\ket{L}$ and $\ket{B}$ are not orthogonal to each other, their normalized superposition is given by
\begin{equation}
\Ket{\psi} = \frac{1}{\sqrt{2(1+\Braket{L|B})}}(\Ket{L}+\Ket{B}).
\label{eqn:initial_state}
\end{equation}
The values of $P(L)$ and $P(B)$ for the equal superposition $\ket{\psi}$ are given by
\begin{align}
\label{eq:over}
P\left(L \right) = \Braket{L|\psi}\!\Braket{\psi|L} = \frac{1+\Braket{L|B}}{2} \nonumber \\
P\left(B \right) = \Braket{B|\psi}\!\Braket{\psi|B} = \frac{1+\Braket{L|B}}{2}
\end{align}
The overlap of the states can be approximated by 
\begin{equation}
\label{eq:overlap}
\Braket{L|B} \simeq \sqrt{\frac{LB}{2\pi\hbar}}.
\end{equation}
As was shown in \cite{hofmann}, a product of $LB\approx0.024 (2\pi \hbar)$ is optimal for the observation of the paradox, corresponding to a situation where about $2.4 \%$ of the photons in $\ket{B}$ will be found in the interval $L$ and vice versa.

The time evolution of the wavefunctions can be approximated by considering that the spatial distribution of $\ket{L}$ is quickly dominated by the diffraction pattern corresponding to the momentum distribution, while the spatial distribution of $\ket{B}$ starts out as a diffraction pattern and hardly evolves at all due to the very low values of momentum in the interval $B$. 
\begin{gather}
%\label{eqn:wf_position}
\Bra{x}U\left(t \right)\Ket{L} \simeq \sqrt{\frac{mL}{2\hbar t}} \: \mathrm{sinc}\left(\frac{mL}{2\hbar t}x \right) \nonumber \\ 
\cdot \:\mathrm{exp}\left[i\left(\frac{m}{2\hbar t}x^2 -\frac{\pi}{4} \right)\right] \nonumber \\
%\label{eqn:wf_momentum}
\label{eqn:wf_pos_mom}
\Bra{x}U\left(t \right)\Ket{B} \simeq \sqrt{\frac{B}{2\pi\hbar}} \: \mathrm{sinc}\left(\frac{Bx}{2\hbar} \right).
\end{gather}
The main difference between the two diffraction patterns is the phase dependence, where the flat phase front of $\ket{B}$ represents $p=0$ and the curved phase front of $\ket{L}$ represents the correlation of position $x(t)$ and momentum $p$. It is also worth noting that the phase of $\ket{L}$ at $x(t)=0$ is approximately $-\pi/4$. This is the Gouy phase of the beam in the limit of $z \to \infty$, as obtained from the  evolution of the initial state given in Eq.(\ref{eq:wave}). It appears as a phase difference of $\pi/4$ at $x(t)=0$ in the constructive interference between $\ket{L}$ and $\ket{B}$ that maximizes the sum of the probabilities $P(L)$ and $P(B)$.

The wavefunction $\bra{x} U(t_M)\Ket\psi$ is the time-evolved wavefunction of the superposition $\ket{\psi}$ at time $t=t_M$. From this wavefunction, we can obtain the probability $P(M)$ by integration over the corresponding probability density,
\begin{equation}
P(M) = \int_{-M/2}^{M/2}dx\Braket{x|U(t_M)|\psi}\!\Braket{\psi|U^{\dagger}(t_M)|x}.
\end{equation}
At a time $t_M$ with $Bt_M/m=L$, the wavefunctions of $\ket{L}$ and $\ket{B}$ have the same spatial envelope and their interference is decided by the quadratic dependence of the phase difference on position $x$. Even though the interference effect increases the probability density close to $x=0$, the value of $P(M)$ is not much higher than the rather low value of $2 |\braket{L|B}|^2$ observed for $\ket{L}$ or for $\ket{B}$. On the other hand, constructive interference increases the value of $P(L)+P(B)-1$ from its value of $|\braket{L|B}|^2$ for $\ket{L}$ and $\ket{B}$ to a much higher value of $\braket{L|B}$ as shown in Eq.(\ref{eq:over}). The result is a violation of the particle propagation inequality with a defect probability of around $7\%$. 

The particle propagation paradox demonstrates that quantum particles do not move in a straight line by identifying an interference effect between a position state $\ket{L}$ and a momentum state $\ket{B}$. This interference effect changes the correlations between position and momentum in a way that cannot be expressed by a positive joint probability density over position and momentum. However, it is possible to express the phase space statistics by a non-positive Wigner function. It is therefore interesting to consider how the probabilities that violate Newton's first law relate to the non-positive values of the corresponding Wigner function.

\subsection{Relation between inequality violation and Wigner function negativity}
The Wigner function can be derived from the time dependent probability distribution of position by assuming that every phase space point $(x,p)$ represents motion in a straight line according to Newton's first law. Any violation of Newton's first law must therefore correspond to negative values of the Wigner function. Oppositely, no state can violate the statistical expectations of Newton's first law if its Wigner function is positive. Here, we identify the phase space regions where the interference of position and momentum results in Wigner function negativity. 

Using the Wigner function $W(x,p)$, the probabilities $P(L)$ and $P(B)$ are given by the integral over phase space regions defined by the position and momentum intervals,
\begin{align}
P(L) &= \int_{-\infty} ^{\infty}dp\int_{-L/2} ^{L/2}dx\:W(x,p) \nonumber \\
P(B) &= \int_{-B/2} ^{B/2}dp\int_{-\infty} ^{\infty}dx\:W(x,p) \nonumber \\
P(M) &= \int_{-\infty}^{\infty}dp \int_{-L-p t_M/m}^{L-p t_M/m} dx\:W(x,p)
\end{align}
These are all marginal probabilities of the Wigner function, since they all include an infinite integral along a specific phase space direction.  

\begin{figure}[H]
    \centering
    \includegraphics[width=0.9\linewidth]{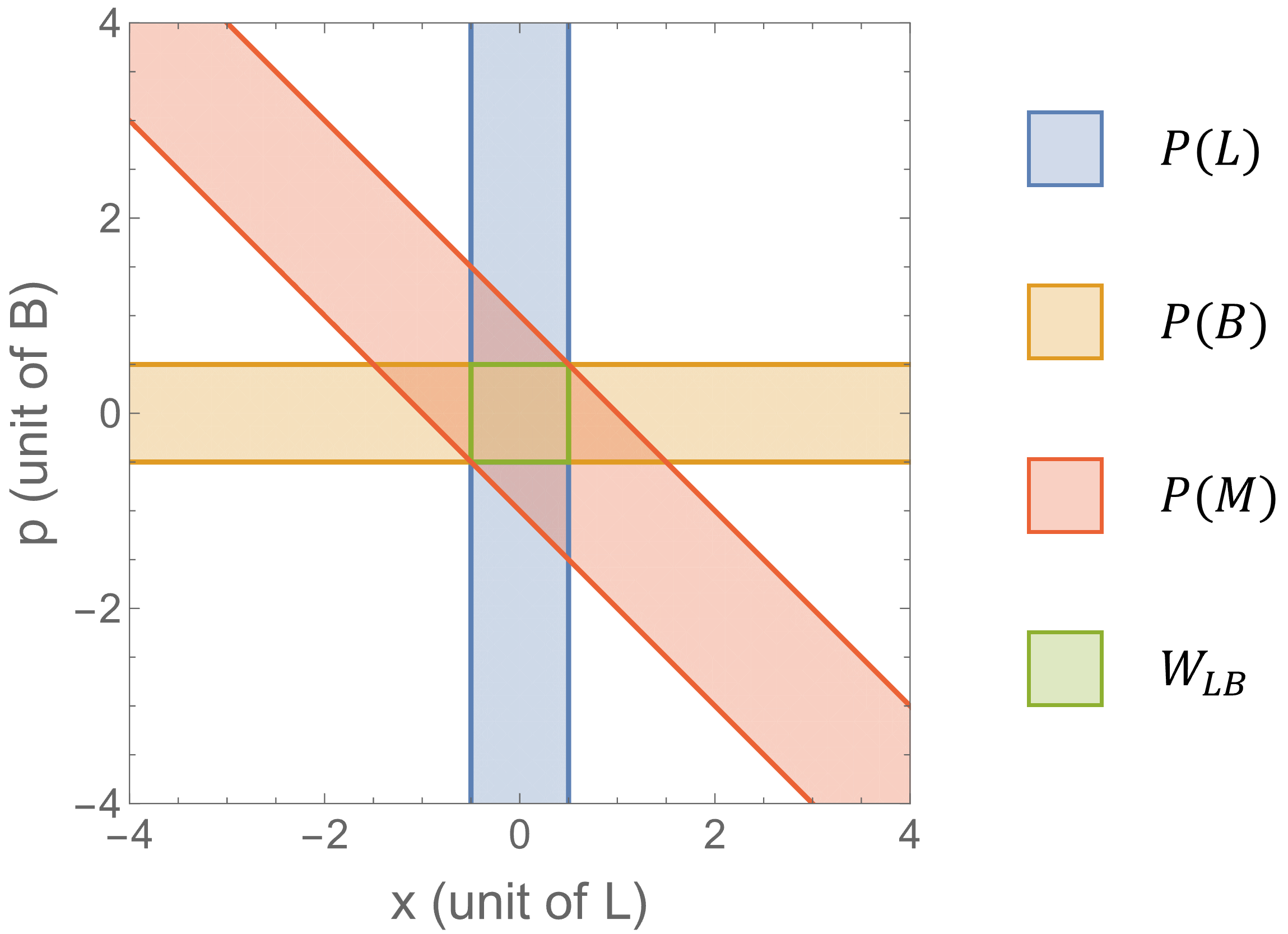}
    \captionsetup{justification=raggedright,singlelinecheck=false}
    \caption{
    Integration region in phase space for the Wigner function.
    }
    \label{fig:Wigner_region}
\end{figure}

To investigate the relation between the inequality violation and negative values of the Wigner function, it is convenient to evaluate the integral of the Wigner function for the cross-shaped region that is either inside the intervals of position $L$ or the interval of momentum $B$. In mathematical terms, this region can be defined as $\Omega = \{(x,p)\:|\:|x| \leq L/2 \:\:\mathrm{or}\:\:|p| \leq B/2\}$ and the integral is given by  
\begin{equation}
\label{eqn:def_W_in}
W_{\mathrm{in}} = \int_{\Omega}dxdp\:W(x,p) 
\end{equation}
The quasi probability $W_{\mathrm{in}}$ is related to the marginal probabilities $P(L)$ and $P(B)$ by
\begin{equation}
\label{eqn:solve_W_in}
W_{\mathrm{in}} = P(L)+P(B)-W_{LB}, 
\end{equation}
where
\begin{equation}
\label{eqn:Wigner_partial_int}
W_{LB} = \int_{-L/2} ^{L/2}dx\int_{-B/2} ^{B/2}dp\:W(x,p).
\end{equation}
is the integral over the region that is both in the interval $L$ and in the interval $B$. $W_{LB}$ contributes to all of the marginal probabilities $P(L)$, $P(B)$, and $P(M)$. Fig.\ref{fig:Wigner_region} illustrates the regions contributing to the different marginal probabilities and indicates the intersection of the regions that defines the integral $W_{LB}$. 

Since the Wigner function is normalized so that the total integral is one, the integral $W_{\mathrm{out}}$ over the region outside of the cross-shaped region including all points in either of the intervals is given by 
\begin{equation}
    W_\mathrm{out} = 1- W_{\mathrm{in}}.
\end{equation}
To express the inequality entirely in terms of integrals of the Wigner function over different regions of phase space, we also define $W_{\mathrm{diag}}$ as the integral over the region inside the interval $M$, excluding only the region jointly in $L$ and $B$,
\begin{equation}
 W_{\mathrm{diag}} = P(M)-W_{LB}.
\end{equation}
Using these definitions, the violation of our inequality can be expressed as
\begin{equation}
\begin{split}
    P_{\mathrm{defect}} &= P(L)+P(B)-1-P(M)\\
    &= (W_{\mathrm{in}}+W_{LB}-1)-(W_{LB}+W_{\mathrm{diag}})\\
    &= -(W_{\mathrm{out}}+W_{\mathrm{diag}}).
\end{split}
\label{eq:defect}
\end{equation}
This relation shows that a positive defect probability necessarily requires negative values of the Wigner function, specifically in the regions associated with $W_{\mathrm{out}}$ and $W_{\mathrm{diag}}$, which are mostly outside of the intervals $L$ and $B$. 

It is also worth noting that the Wigner function integral $W_{LB}$ corresponds to a joint probability of $L$ and $B$, as suggested by Eq.(\ref{eqn:solve_W_in}). However, the Wigner function cannot exceed an upper bound of $1/(\pi\hbar)$, limiting the quasi probability $W_{LB}$ to
\begin{equation}
\label{eq:WLBmax}
    W_{LB} \leq \frac{LB}{\pi\hbar}.
\end{equation}
It is possible that the minimal joint probability of $L$ and $B$ given by $P(L)+P(B)-1$ exceeds this limit. The additional probability must then originate from $W_{\mathrm{in}}>1$ and $W_{\mathrm{out}}<0$. Specifically, the inequality for $W_{\mathrm{out}}$ reads
\begin{equation}
    W_{out} \leq 1-P(L)-P(B)+\frac{LB}{\pi \hbar}.
\label{eq:negative_wigner}
\end{equation}
In this relation, the quantum mechanical limit of $LB/(\pi \hbar)$ replaces the probability $P(M)$ as upper bound of the joint probability $P(L \;\mathrm{AND}\; B)$ for positive quasi-probabilities. If the minimal joint probability given by $P(L)+P(B)-1$ exceeds this quantum mechanical limit, the Wigner function outside of the cross-shaped region defined by the intervals $L$ and $B$ has a negative integral $W_{\mathrm{out}}$. In the following, we will apply this relation to demonstrate the negativity of the Wigner function from the experimentally observed values of $P(L)$ and $P(B)$ and identify the phase space region outside of the intervals $L$ and $B$ as the location of this negativity. 

%%%=================================================

\section{\label{sec:level3}Experiment} 
\begin{figure*}[t]
    \centering
    \includegraphics[width=0.66\textwidth]{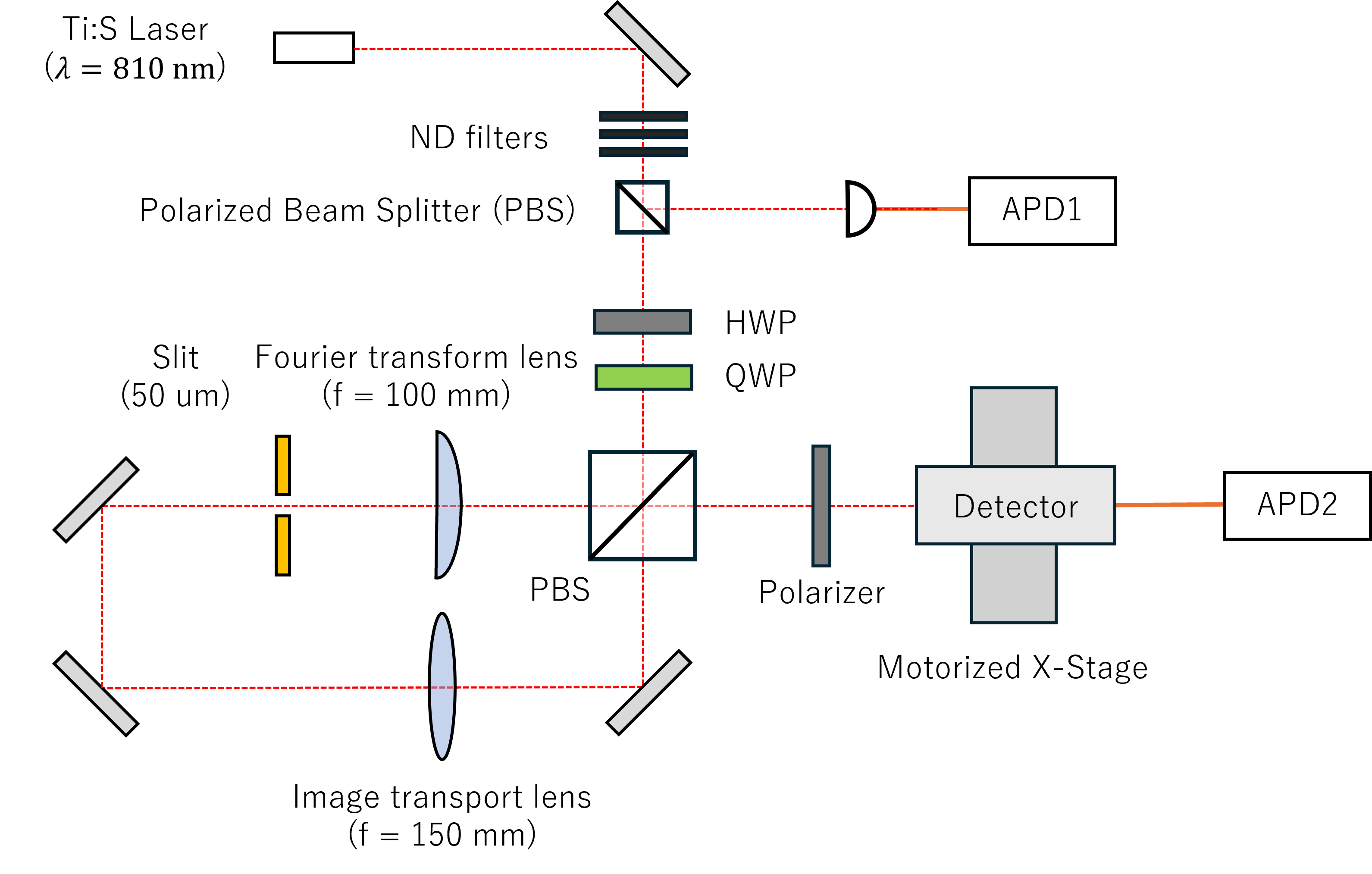}
    \captionsetup{justification=raggedright,singlelinecheck=false}
    \caption{
    Experimental Setup. The initial light with $\lambda = 810 \:\mathrm{nm}$ from the Ti:Sapphire laser was attenuated to the single photon level using ND filters and its stability was monitored by an avalanche photo diode module APD1. The input polarization used for fine-turning of amplitude and phase of the superposition state was adjusted by a half wave plate and a quarter wave plate. As a result, a superposition of the $\ket{L}$ state and the $\ket{B}$ state was prepared by the interference at the exit of the interferometer. The transverse photon distribution was measured with a detector system mounted on a movable stage connected to another avalanche photo diode module APD2 via a multi-mode fiber.
    }  
    \label{fig:setup}
\end{figure*}

We have investigated the distribution of transverse position and momentum of photons exiting from a Sagnac interferometer designed to prepare a superposition of the position state $\ket{L}$ and the momentum state $\ket{B}$. 
In this experiment, the stability of the initial superposition state is very important. It should be noted that the spatial modes of the two components $\Ket{L}$ and $\Ket{B}$ are completely different from each other. The visibility of interference in the total transmission of the interferometer is therefore very low. For the stabilization of an optical interferometer, such low visibility makes feedback control very difficult. The inherent stability of the Sagnac interferometer is therefore essential.\\
\begin{figure}[b]
    \centering
    \includegraphics[width=0.7\linewidth]{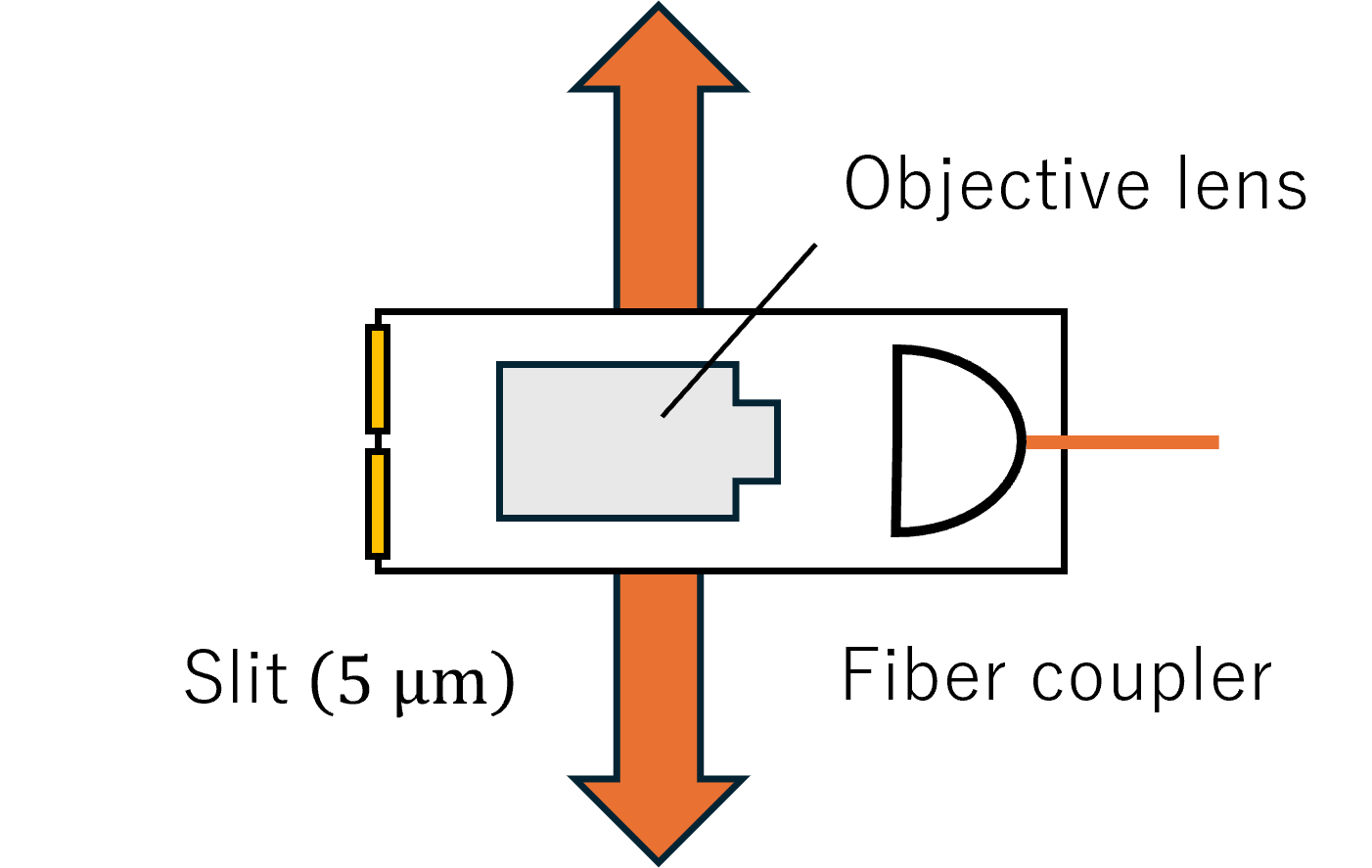}
    \captionsetup{justification=raggedright,singlelinecheck=false}
    \caption{
    Detection system to measure the transverse distribution. The photons pass through a slit of $5 \:\mathrm{\mu m}$ coupled to a multi-mode fiber by an objective lens. This system was mounted on an automated X-stage to be moved in increments of $5 \:\mathrm{\mu m}$ in a transverse direction. The count rate total of background counts from stray photons and dark noise in each position was less than 190/s, measured over a period of 2 minutes at each position, corresponding to a total of 40 hours of background measurements. 
    }
    \label{fig:comp_system}
\end{figure}

\begin{figure*}[t]
  \centering

  \hfill
  \begin{subfigure}[b]{0.49\textwidth}
    \centering
    \includegraphics[width=\linewidth]{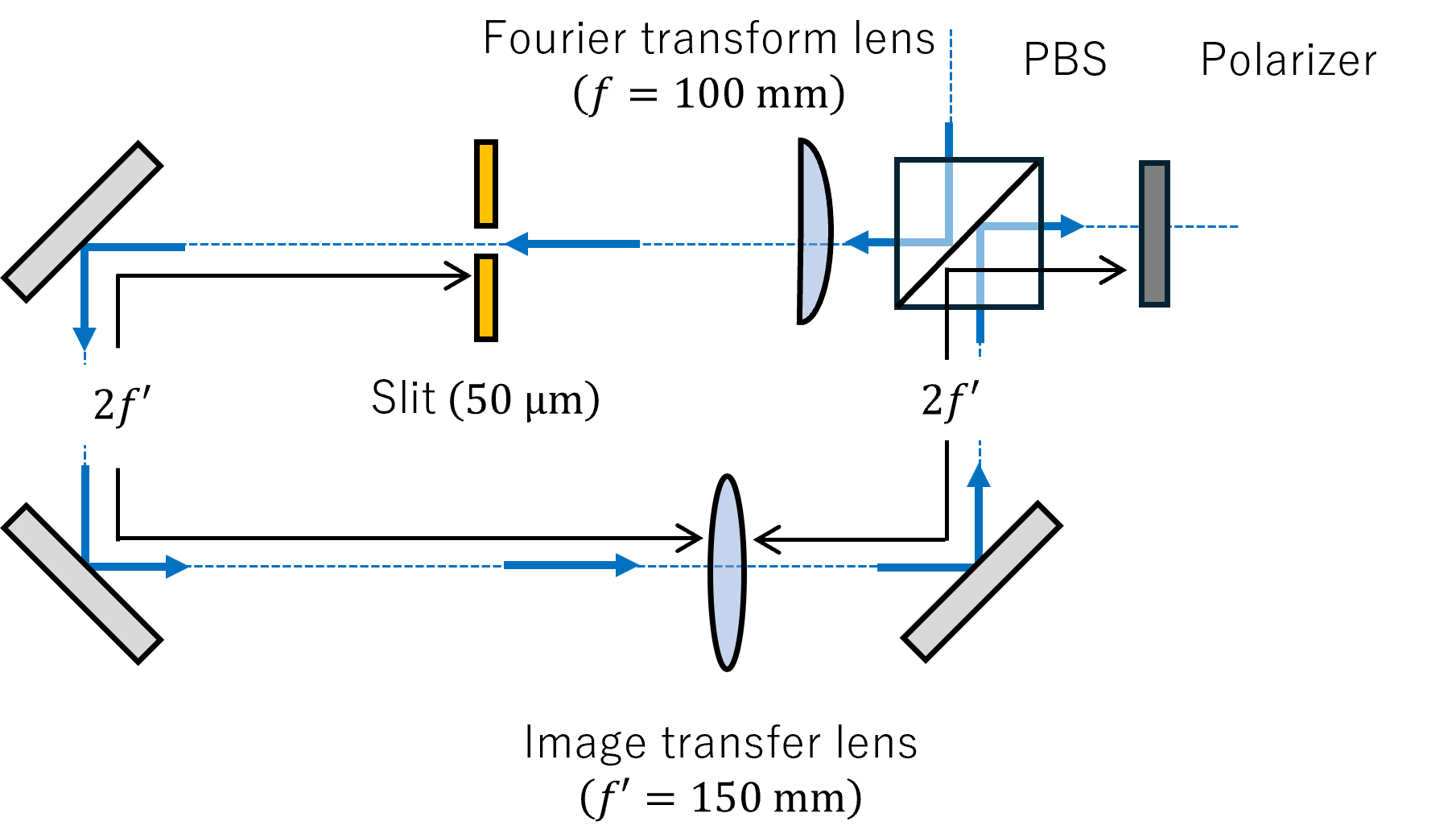}
    \caption{}
    \label{fig:anticlockwise}
  \end{subfigure}
  \hfill
  \begin{subfigure}[b]{0.49\textwidth}
    \centering
    \includegraphics[width=\linewidth]{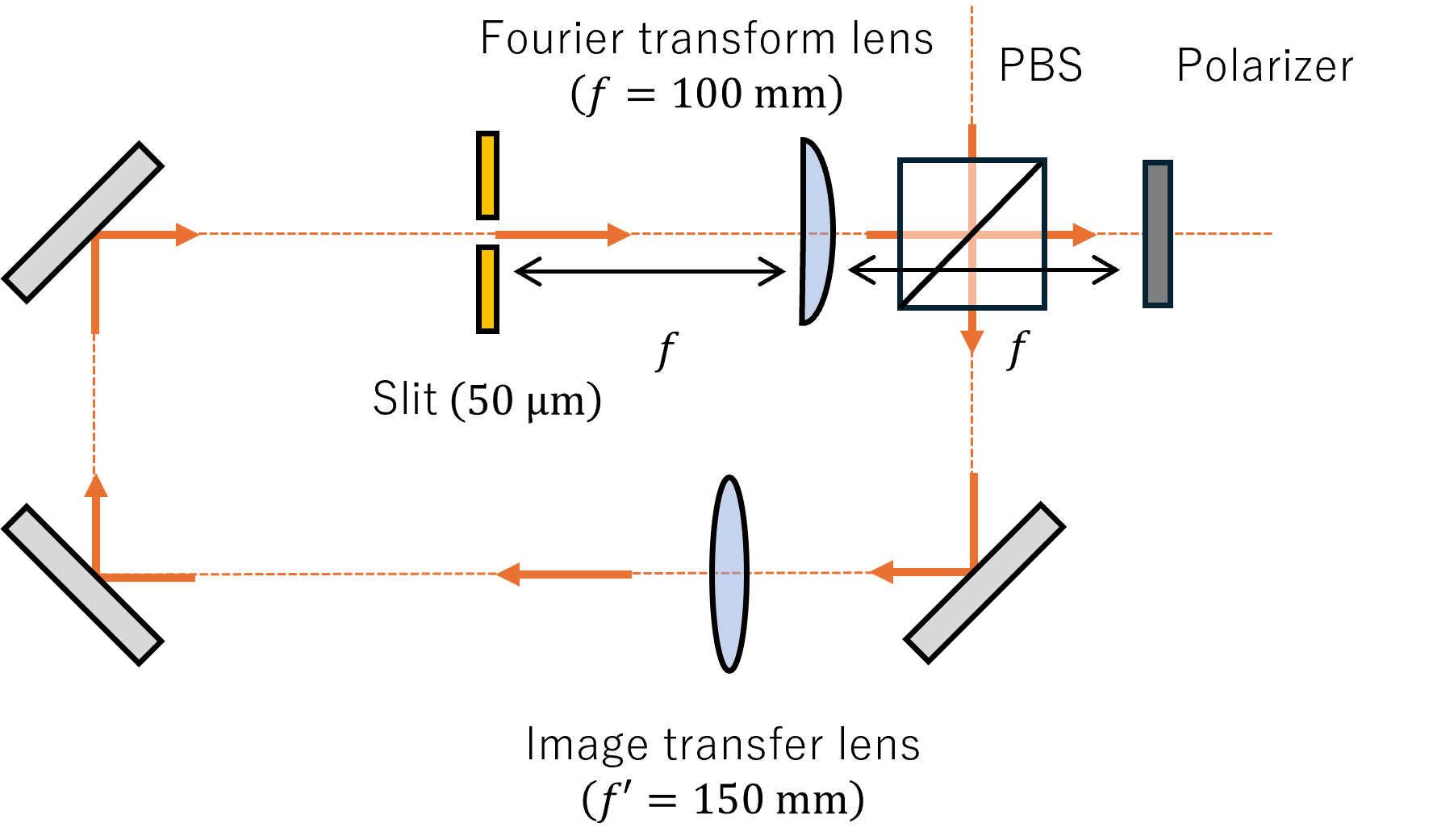}
    \caption{}
    \label{fig:clockwise}
  \end{subfigure}   \captionsetup{justification=raggedright,singlelinecheck=false}
  \caption{
  Preparation of $\ket{L}$ and $\ket{B}$ states in the Sagnac Interferometer. % (a): Figure of our Sagnac interferometer   
  (a): Preparation of the state $\ket{L}$. The rectangular distribution just after the slit is transferred by the biconvex lens. 
  (b): Preparation of the state $\ket{B}$. The distance between the slit and the Fourier transform lens was adjusted to be equal to its focal length. The rectangular distribution in position space was thus transformed into the rectangular distribution in momentum space.
  }
  \label{fig:sagnac_path}
\end{figure*}

Fig. \ref{fig:setup} shows the whole experimental setup. We used a Ti:sapphire laser with $\lambda = 810 \;\mathrm{nm}$ as a photon source. The input intensity was sufficiently attenuated with ND filters with transmittance $2.8\times10^{-10}$ to allow reliable single photon detection in the output.
The input photons passed through a polarization beam splitter (PBS) and their polarization state was controlled with a Half-wave-plate (HWP) and a Quarter-wave-plate (QWP) located behind the PBS. After splitting the two paths at a large size PBS (5cm cubic), the anticlockwise and the clockwise paths were used to prepare the $\Ket{L}$ state and the $\Ket{B}$ state, respectively, and these two paths were then interfered at the same PBS. Since the polarization state in the anticlockwise path is vertical (V) and the one in the clockwise path is horizontal (H), the appropriate selection of polarization with a polarizer placed behind the PBS prepares a superposition state $\ket{\psi}$. At a certain distance from the PBS, we measured the probability distributions of initial position, of initial momentum, and of intermediate position (effective time $t_M$). In each measurement, the photons entered a detection system where a slit of $5 \:\mathrm{\mu m}$ was coupled to a multi-mode fiber via an objective lens as shown in Fig. \ref{fig:comp_system}. The multi-mode fiber was connected to an avalanche photo diode module (APD2) specifically designed for connection to a multi-mode fiber. The detection system was mounted on an automated X-stage to be moved in increments of $5 \:\mathrm{\mu m}$ in a transverse direction. The detection system was moved in the transverse direction in steps, and photons were counted at each position. To monitor the intensity of the initial photons, we placed another APD (APD1) in the reflection port of the upstream PBS and recorded the stability of the intensity. The intensity data can be used for removing the influence of the intensity instability. Furthermore, background photons and dark noise were counted by blocking the input photons upstream. The ON/OFF remote control for inputting the photons was realized by a shutter controlling the entry of the light. The APDs used are SPCM-AQR-1410117 manufactured by PerkinElmer for monitoring the input light and SPCM-AQRH-14-FC24360 manufactured by EXCELITAS for measuring the distributions. Their typical detection efficiencies at 810 nm are 54 \% and 62 \%, and their typical dead times are 50.7 ns and 29.3 ns, respectively.  Photon events were measured over a period of 2 minutes for each step to accumulate sufficient number of counts, once without input to obtain background counts, and once with input to determine the pattern. For the entire distribution, this corresponds to 80 hours of measuring time, 40 hours for the background and 40 hours for the signal. The count rate total of background counts was less than 190/s for all measurements. The highest count rate including background is less than $5500\:/\mathrm{s}$. Based on this value, we estimated the probability that a second photon arrives at the detecter during its dead time. Assuming Poissonian statistics, this probability is given by $5500\:/\mathrm{s} \times 30 \:\mathrm{ns} = 1.65 \times 10^{-4}$ or $0.0165 \:\%$. Even at the highest observed count rates, the fraction of photons arriving during the dead time of another detection event is only about 1 in 6000. We can therefore confirm that all of the count rates correspond to single photon probabilities.

The specific way for preparing both of $\ket{L}$ and $\ket{B}$ states is shown in Fig. \ref{fig:sagnac_path}. The $\ket{L}$ state was created by transferring the spatial mode at the slit of the width $L$ with a biconvex lens as shown in Fig. \ref{fig:sagnac_path}(a). On the other hands, Fig. \ref{fig:sagnac_path}(b) shows how the $\Ket{B}$ state was created by a Fourier Transform of the spatially-rectangular mode of the width $d$ with a plano-convex lens of a focal length $f$. Placing this lens at the focal length away from the slit produced the rectangular mode in momentum space with the width of $B= hd/\lambda f$ at a focal plane further away.  

\begin{figure}[h]
    \centering
    \includegraphics[width=0.9\linewidth]{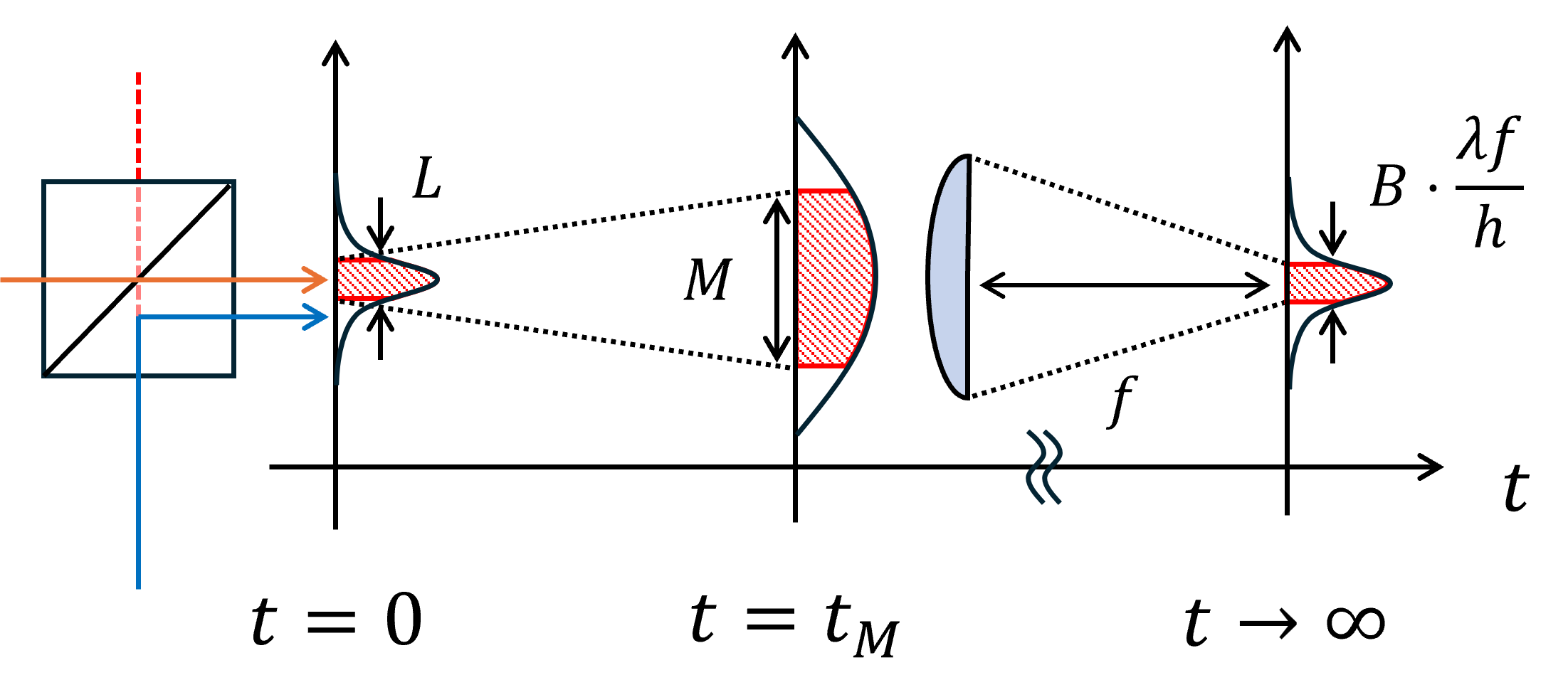}
    \captionsetup{justification=raggedright,singlelinecheck=false}
    \caption{
    Measurements of the spatial distribution at $t=0$, $t=t_M$, and $t\rightarrow \infty$. We experimentally defined the position of the image plane of $\Ket{L}$ as $t=0$, which corresponds to the measurement of initial position. The measurement at $t=t_M$ corresponds to the plane at $z=ct_M$ downstream from the initial plane, which matches the focal length of the Fourier transform lens in the interferometer at $f=100\:\mathrm{mm}$. The measurement of $t\rightarrow \infty$ can be achieved by using an additional lens to implement a Fourier transform. The transverse momentum is given by the angle of propagation at $t=0$ which determines the transverse position at infinite distance. The role of the convex lens is to map the angle of propagation in the $t=0$ plane onto a corresponding position in the focal plane. Thus, a measurement performed in the focal plane of the convex lens is equivalent to the measurement of the scaled momentum distribution, corresponding to an observation of the distribution at $t\rightarrow\infty$.
    }
    \label{fig:measurement}
\end{figure}

Fig \ref{fig:measurement} illustrates the different measurement settings for the distributions of initial position, initial momentum, and intermediate position. The initial position is observed in a plane at $z=0$ corresponding to the initial time of $t=0$ as indicated in Fig. \ref{fig:concept}. The position distribution of the initial superposition state $\Ket{\psi}$ was measured by directly placing the detector at the plane of $t=0$. The momentum distribution at $t=0$ is obtained by Fourier transforming spatial distribution. We prepared another plano-convex lens with the same model as the one to prepare state $\ket{B}$ in the interferometer and placed $f =100 \:\mathrm{mm}$ downstream from the $t=0$ plane. The rectangular image of state $\ket{B}$ can be used for the alignment of the detector along the optical axis to place the detector at the focal plane of the lens. Finally, the intermediate distribution at $t=t_M = mL/B$ can be obtained at a distance of $ct_M = 2\pi \hbar LB/\lambda$ behind the initial plane, where the paraxial approximation is used to separate longitudinal and transverse motion. Note that the transverse velocity of the photon can be approximated by $p_x c \lambda/(2 \pi\hbar)$, where $p_x$ is the transverse momentum. 

Experimental parameters were selected based on the theoretical conditions for high values of $P_{defect}$ while keeping $t=t_{M}$ at a reasonable distance for the detector setting of the intermediate measurement. We chose the condition of $d = 50 \:\mathrm{\mu m}$, $f=100 \:\mathrm{mm}$, which corresponds to $LB/(2 \pi \hbar) = 0.0309$ and $ct_M =100\:\mathrm{mm}$. Using these parameters, the $t=0$ plane was expected to be located $50\:\mathrm{mm}$ behind the large size PBS at the exit of the interferometer. 

The phase difference between $\Ket{L}$ and $\Ket{B}$ can be tuned to zero by monitoring the center of the position distribution at the intermediate setting of $z=ct_M$ where the contributions from the two components are approximately equal and the local phase difference should be $\pi/4$ for maximal constructive interference. The correct phase adjustment is obtained by achieving a phase of $-\pi/4$ for V polarized light relative to H polarized light at $x\simeq 0$. The amplitudes can be adjusted by setting a corresponding input polarization using the HWP and the QWP shown in Fig. \ref{fig:setup}.

%%%=========================================================

\section{\label{sec:level4}Results and Discussion} 
\subsection{Position and momentum at $t=0$}
\begin{figure*}[hbtp]
  \centering
  \includegraphics[width=0.98\textwidth]{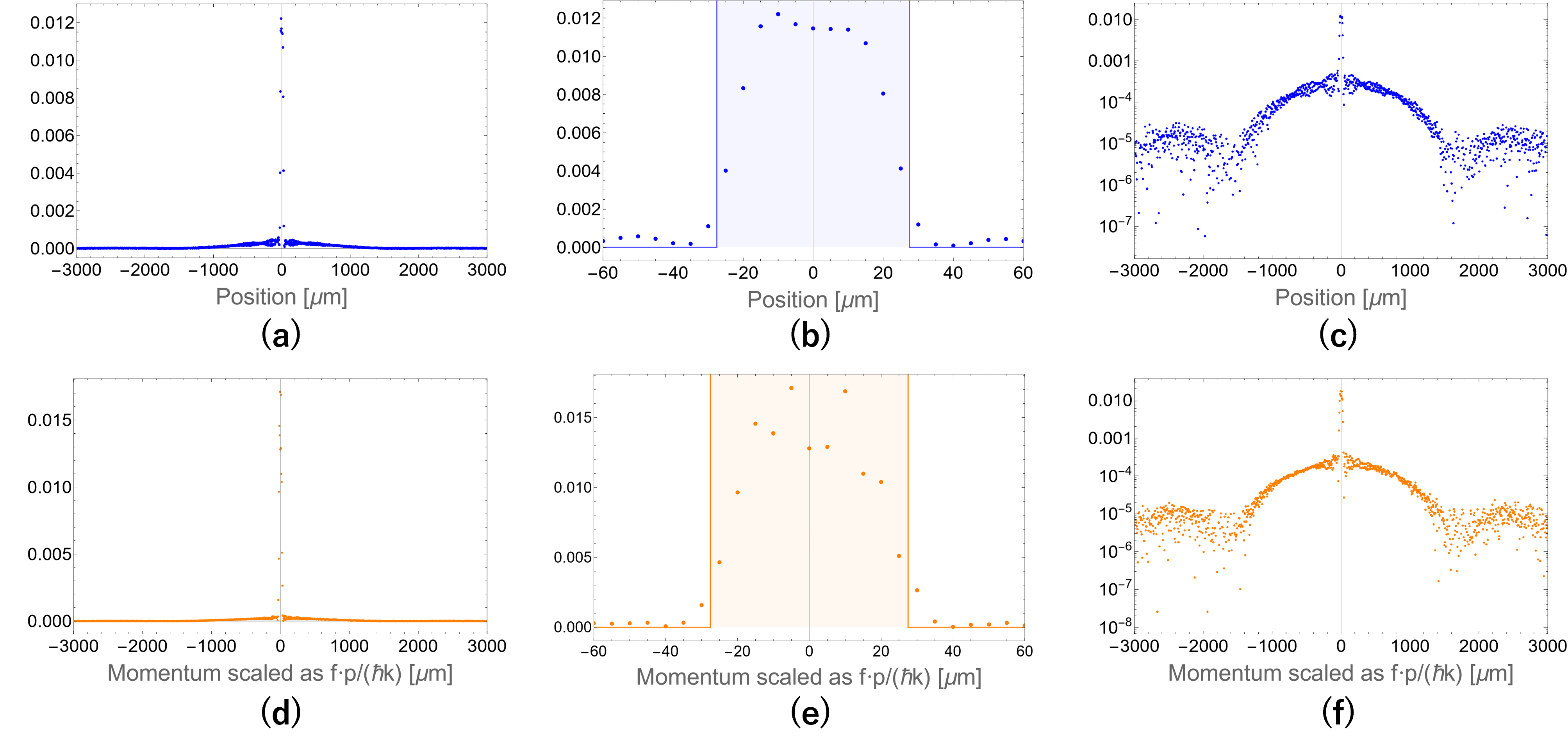}
  \captionsetup{justification=raggedright,singlelinecheck=false}
  \caption{
  Experimental results for the distributions of position and momentum at time $t=0$. To accumulate a sufficient number of counts, we measured over a period of 2 minutes for each step, once for the background and once for the signal. This corresponds to 80 hours of measurement time for the entire distribution. Background counts were subtracted, and the remaining counts were divided by the input signal to compensate for any instability in the intensity of the  laser during the measurement. The result was then normalized based on the total of all results obtained over the full range of positions. We thus obtain a probability density in units of $\mathrm{\mu m^{-1}}$. (a) shows the position distribution over a range of 6 mm. (b) shows details of the distribution over a range of $120 \mu m$ around the interval of $L=55 \mu m$ (shaded area). (c) shows a logarithmic plot of the probability density to highlight the diffraction pattern associated with $\ket{B}$. (d) to (f) show corresponding graphs for the momentum distribution, where the momentum is scaled to $f p/(\hbar k)$. Using this scale, the interval $B$ in (e) is also $55 \mu m$ wide, and the values of the densities are comparable as well. 
  }
  \label{fig:6images}
\end{figure*}

We first consider the measurement results for the distribution of position and momentum at $t=0$. Fig. \ref{fig:6images} shows the results after background subtraction and division by the input intensity monitored by APD1 to compensate for any instability in the intensity of the laser during the measurement. The results are normalized over the full range of positions to obtain a probability density.
As expected, the distributions have two distinct features. The first is a nearly rectangular peak having a width of $L$ for position and $B$ for momentum. Since the peak is not perfectly rectangular, we have defined an interval of $L=55 \:\mathrm{\mu m}$ for the evaluation of $P(L)$. For better comparison, we have scaled the momentum to $f p/(\hbar k)$, where $B = 55\:\mathrm{\mu m}\:\cdot \hbar k/f$. $k=2 \pi/\lambda$ is the wavenumber and $f=100 \mathrm{mm}$ is the focal length of the lens used in the setup. For the detection intervals of $55 \:\mathrm{\mu m}$, the product of the intervals is given by $LB \simeq 0.037 (2 \pi \hbar)$, which is about 20 \% higher than the result obtained for the slit width of $50 \:\mathrm{\mu m}$ used in the setup. Figs. \ref{fig:6images} (b) and (e) highlight this feature by showing only the probability densities of the position and momentum values close to the intervals $L$ and $B$.

The second feature is the much wider diffraction pattern corresponding to the slit diffraction of $\ket{L}$ in the momentum distribution and the corresponding diffraction pattern of $\ket{B}$ in the position distribution. Since the probability densities of this feature are much lower than those of the nearly rectangular peaks, it is difficult to show both features in the same graph. Figs. \ref{fig:6images} (c) and (f) show the densities on a logarithmic scale. For a slit width of $50 \:\mathrm{\mu m}$, we expect the minima to be at $\pm 1.6 \:\mathrm{mm}$ and this appears to be roughly consistent with the results shown.  

In the measurement of position, only the component $\ket{B}$ contributes significantly to positions $|x|> L/2$. Likewise, only the component $\ket{L}$ contributes significantly to momenta with $|p|>B/2$. Interferences between $\ket{L}$ and $\ket{B}$ should therefore be limited to the intervals $L$ and $B$ in both the position and the momentum measurements. We can express the decomposition into statistical contributions from $\ket{B}$, from $\ket{L}$, and from their interference by expressing the density operator in the form 
\begin{equation}
\begin{split}
    \hat\rho = & w_{B}\Ket{B}\!\Bra{B}+w_{L}\Ket{L}\!\Bra{L} \\[0.2cm]
         &+ w_{\mathrm{inter}} \left(\:\frac{\Ket{L}\!\Bra{B}}{2\Braket{B|L}}+\:\frac{\Ket{B}\!\Bra{L}}{2\Braket{L|B}}\right).
\end{split}
\label{eq:exp_state}
\end{equation}
Here, $w_L$, $w_B$, and $w_{\mathrm{inter}}$ are quasi-probabilities that sum up to one,
\begin{equation}
\label{eq:norm}
    w_{L}+w_{B}+ w_{\mathrm{inter}} = 1.
\end{equation}
The interference effect can be associated with a fraction of $w_{\mathrm{inter}}$ of the photons observed in the experiment, and we can identify the characteristics of this fraction of the observed photons by considering the manner in which the interference term in Eq.(\ref{eq:exp_state}) appears in the experimentally observed statistics. The contributions of the quasi-probabilities to the values of $P(L)$ and $P(B)$ are given by
\begin{align}
\label{eq:quasiw}
P(L) &= w_L +w_{\mathrm{inter}} + w_B |\Braket{B|L}|^2 \nonumber \\
P(B) &= w_B +w_{\mathrm{inter}} + w_L |\Braket{B|L}|^2.
\end{align}
These equations suggest that the photon counts in the intervals $L$ and $B$ can be broken down into three contributions associated with $\ket{L}$, with $\ket{B}$, and with the interference of the two. Here it is important to note that the probability $w_{\mathrm{inter}}$ of the interference term fully contributes to both $P(L)$ and $P(B)$, indicating that the interference term represents photons that are jointly found in $L$ and in $B$. Indeed, we find that
\begin{equation}
P(L)+P(B)-1 = w_{\mathrm{inter}} + (w_L+w_B) |\Braket{B|L}|^2, 
\end{equation}
indicating that the interference effect describes the statistics of photons in both $L$ and $B$. 

From the probability density distributions shown in Fig. \ref{fig:6images} (a) and (d), which are obtained by normalizing the counts after background substraction and compensation of intensity fluctuations, we determine the probabilities by numerically integrating the probability densities over the corresponding intervals highlighted in the enlarged views shown in Figs. 7(b) and 7(e). The values of the probabilities are $P_{\mathrm{exp}}(L) = 0.5249 \:\pm\:0.0007$ and $P_{\mathrm{exp}}(B) = 0.6445 \:\pm\:0.0007$. According to Eq. (\ref{eq:overlap}), we expect the slit width of $50 \:\mathrm{\mu m}$ in the current setup to result in an overlap of $|\braket{B|L}|^2=0.0309$. If this value is correct, we can determine the complete quasi-probability from the values of $P(L)$ and $P(B)$ using
\begin{align}
w_L &= \frac{1-P_{\mathrm{exp}}(B)}{1-|\braket{B|L}|^2} = 0.367 \nonumber \\
w_B &= \frac{1-P_{\mathrm{exp}}(L)}{1-|\braket{B|L}|^2} = 0.490, 
\end{align}
and $w_{\mathrm{inter}}=0.143$ to complete the probability total of one. 

This initial evaluation of the input state relied only on the total number of counts inside and outside of the intervals $L$ and $B$. However, we can obtain additional information by considering the precise shape of the diffraction pattern. Specifically, the contributions of $w_B$ to P(L) and of $w_L$ to P(B) can be identified by interpolating the values of the diffraction patterns inside the intervals $L$ and $B$, noting that the integral over this interpolation is a good approximation of the overlap between the underlying states,
\begin{align}
w_B |\Braket{B|L}|^2 &\simeq \int_{-L/2}^{L/2} w_{B}\Braket{x|B}\!\Braket{B|x}dx \\
 w_{L}|\Braket{B|L}|^2 &\simeq \int_{-B/2}^{B/2} w_{L}\Braket{p|L}\!\Braket{L|p}dp.
\label{eqn:LtoB}
\end{align}
Although the experimental data does not allow us to isolate the above probability densities in the intervals $L$ and $B$, it is possible to confirm that the diffraction pattern outside of that region is described by a sinc function, and the interpolation of this sinc function will then provide an estimate of the contributions of $\ket{B}$ to $P(L)$ and of $\ket{L}$ to $P(B)$.

\begin{figure*}[htbp]
  \centering
  \par\medskip
  \vspace{5pt}
  \hfill
  \begin{subfigure}[b]{0.49\textwidth}
    \centering
    \includegraphics[width=\linewidth]{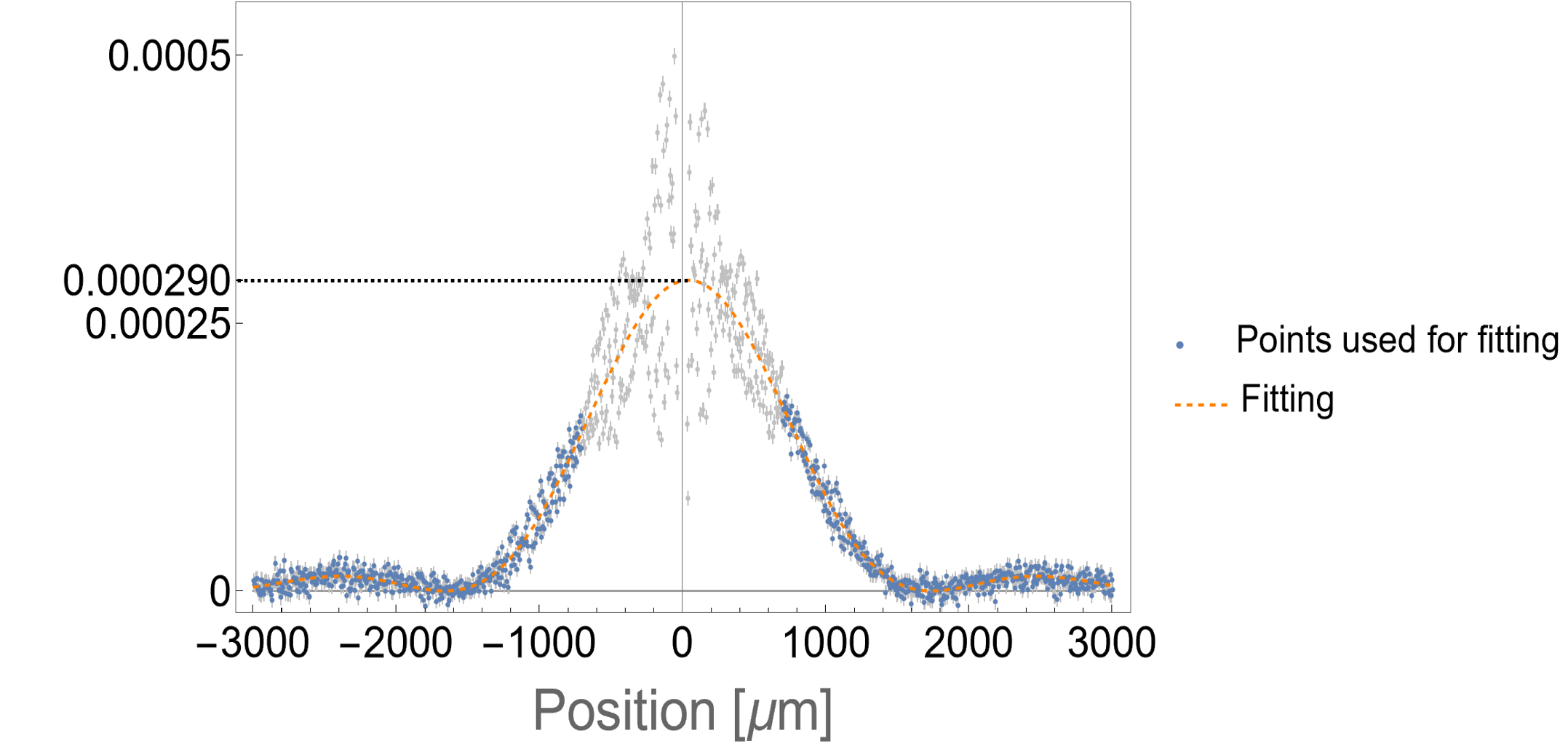}
    \caption{}
  \end{subfigure}
  \hfill
  \begin{subfigure}[b]{0.49\textwidth}
    \centering
    \includegraphics[width=\linewidth]{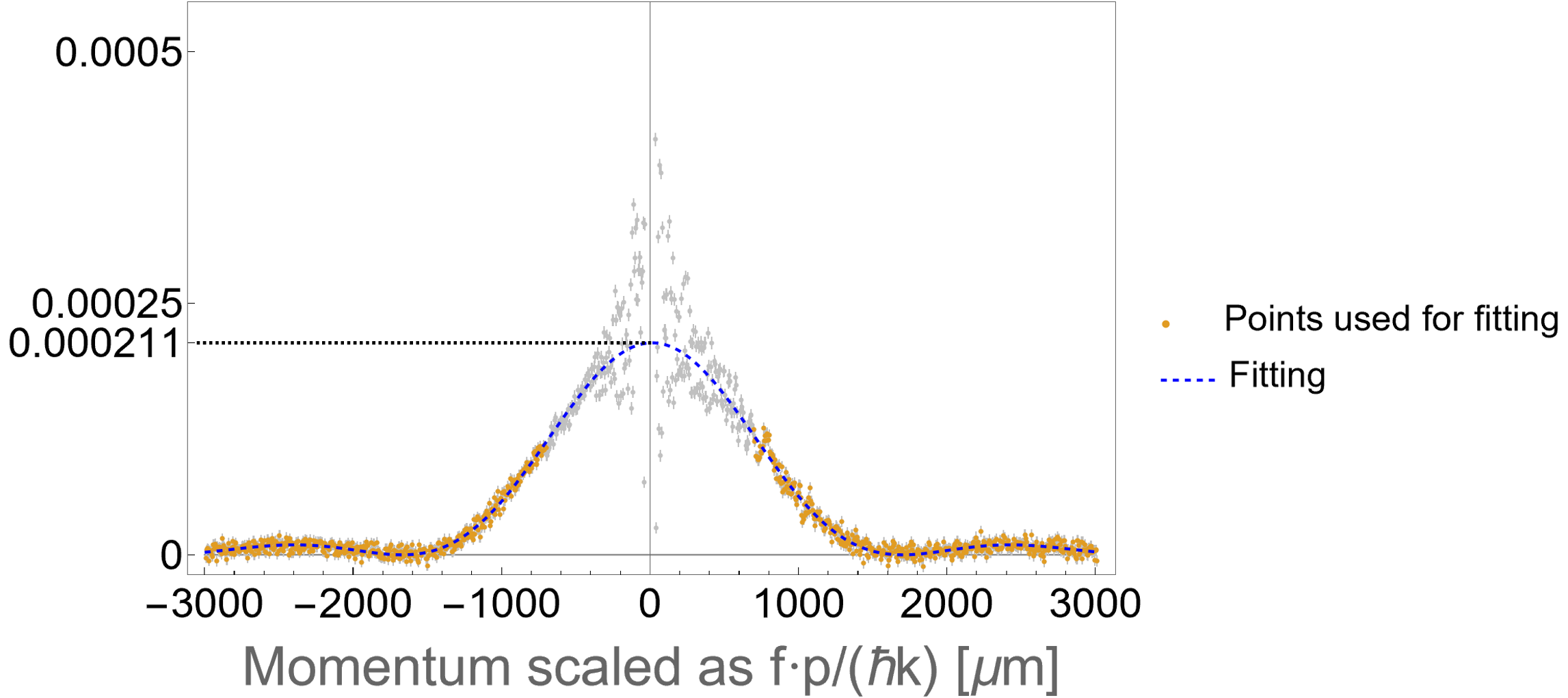}
    \caption{}
  \end{subfigure}   \captionsetup{justification=raggedright,singlelinecheck=false}
  \caption{
  Fits of the diffraction patterns. The vertical axis shows the probability density in unit of $\mathrm{\mu m^{-1}}$. (a) shows the position distribution and (b) shows the momentum distribution scaled to $f p/(\hbar k)$. The fit is complicated by interference effects between $\ket{L}$ and $\ket{B}$ outside of the intervals of $L$ and $B$. These interference effects occur because the measurement is not located exactly in the image plane of the slit. 
  }
  \label{fig:fittail}
\end{figure*}

To confirm the size and shape of the sinc function contribution in the position and momentum distributions, we removed the central part from each distribution and fitted a sinc function to the remaining data. The fitting parameters of the sinc functions are the location of the first minimum which should occur at $1620 \:\mathrm{\mu m}$ for a slit width of $50 \:\mathrm{\mu m}$, and the height which can be used to determine the contributions of $w_B$ in $P(L)$ and of $w_L$ in $P(B)$.
The results of the fits are shown in Fig. \ref{fig:fittail}. Near the center, the densities both show an oscillation indicating an additional interference effect between $\ket{L}$ and $\ket{B}$.  This interference effect involves non-vanishing amplitudes of $\ket{L}$ outside of the interval $L$ and non-vanishing amplitudes of $\ket{B}$ outside of the interval $B$. The most likely origin is the onset of diffraction caused by small deviation of the measurement plane $z$ from the image plane of the slit. To simplify the fit, the window of omitted data was widened and only the colored points in Fig. \ref{fig:fittail} were included in the fit. The location of the first minimum in the fit of the position distribution was $1700 \pm 8 \:\mathrm{\mu m}$ and the location of the first minimum in the fit of the scaled momentum distribution was $1686 \pm 6 \:\mathrm{\mu m}$. These values are about 4 \% higher than the expected values, suggesting that the actual slit width might be closer to $48 \:\mathrm{\mu m}$. The actual overlap of the states would then be reduced to $|\braket{B|L}|^2=0.0285$. The fits can also be used to obtain more reliable values for the quasi-probabilities in Eq.(\ref{eq:exp_state}). The results obtained from the integrals of the squared sinc functions are $w_B = 0.493\pm0.006$ and $w_L = 0.355\pm0.0004$ respectively. Normalization then results in a value of $w_{\mathrm{inter}} = 0.152\pm0.007$ for the quasi-probability of interference. 

The results of the analysis allow us to determine the contributions of $w_B$ to $P(L)$ and of $w_L$ to $P(B)$. These contributions are found to be $w_B|\braket{B|L}|^2 = 0.0159\pm0.002$ and $w_L|\braket{B|L}|^2=0.0116\pm0.0014$. It should be noted that these contributions are actually found by integrating the distribution over the intervals $L$ and $B$. Since these intervals are significantly wider than the states $\ket{L}$ and $\ket{B}$, the values of $|\braket{B|L}|^2$ will systematically overestimate the actual overlap. Using the results for the quasi-probabilities derived from the same set of fits, we obtain two values for the overlap $\left | \Braket{L|B} \right |^{2}$. For the position distribution, we obtain ($0.032\pm0.005$) and for the momentum distribution, we obtain ($0.033\pm0.004$). The difference between the two results is well within the error margins. As expected, the value is higher than the more reliable one obtained from the location of the minima, where the increase can be explained by the ratio of the detection interval of $55 \:\mathrm{\mu m}$ and the effective slit width of $48 \:\mathrm{\mu m}$. Using only the values obtained from the two diffraction patterns, $P(L)$ and $P(B)$ can be calculated as follows,
\begin{align}
P_{\mathrm{cal}}(L) &= w_L + w_{\mathrm{inter}} + w_B |\Braket{B|L}|^2 \nonumber \\
&=0.523\pm0.009 \nonumber \\
P_{\mathrm{cal}}(B) &= w_B + w_{\mathrm{inter}} + w_L |\Braket{B|L}|^2 \nonumber \\
&=0.657\pm0.009.
\end{align}
These results are roughly consistent with the direct measurement results of $P_{\mathrm{exp}}(L)$ and $P_{\mathrm{exp}}(B)$, although the value of $P_{\mathrm{cal}}(B)$ is slightly higher. 

For the experimental observation of the particle propagation paradox, only the directly observed experimental values of $P_{\mathrm{exp}}(L)$ and $P_{\mathrm{exp}}(B)$ are relevant. The purpose of the quasi-probabilities is to explain the origin of the inequality violation and to identify the role of quantum interference in the propagation of particles in free space.
The minimal joint probability of $L$ and $B$ for the experimental values of $P_{\mathrm{exp}}(L)$ and $P_{\mathrm{exp}}(B)$ is given by $P_{\mathrm{exp}}(L)+P_{\mathrm{exp}}(B)-1 = 0.1694 \pm 0.0010$. According to the fits, the interference term contributes a value of $w_{\mathrm{inter}}=0.152$. Most of the photons in both $L$ and $B$ appear to originate from the interference term. It seems reasonable to conclude that interference between position and momentum implements a joint control of both phase space coordinates. In the following, we shall explore whether this joint control applies to the time evolution of position as well.

\subsection{Position at $t=t_M$}

\begin{figure*}[htbp]
  \centering
  \includegraphics[width=0.8\textwidth]{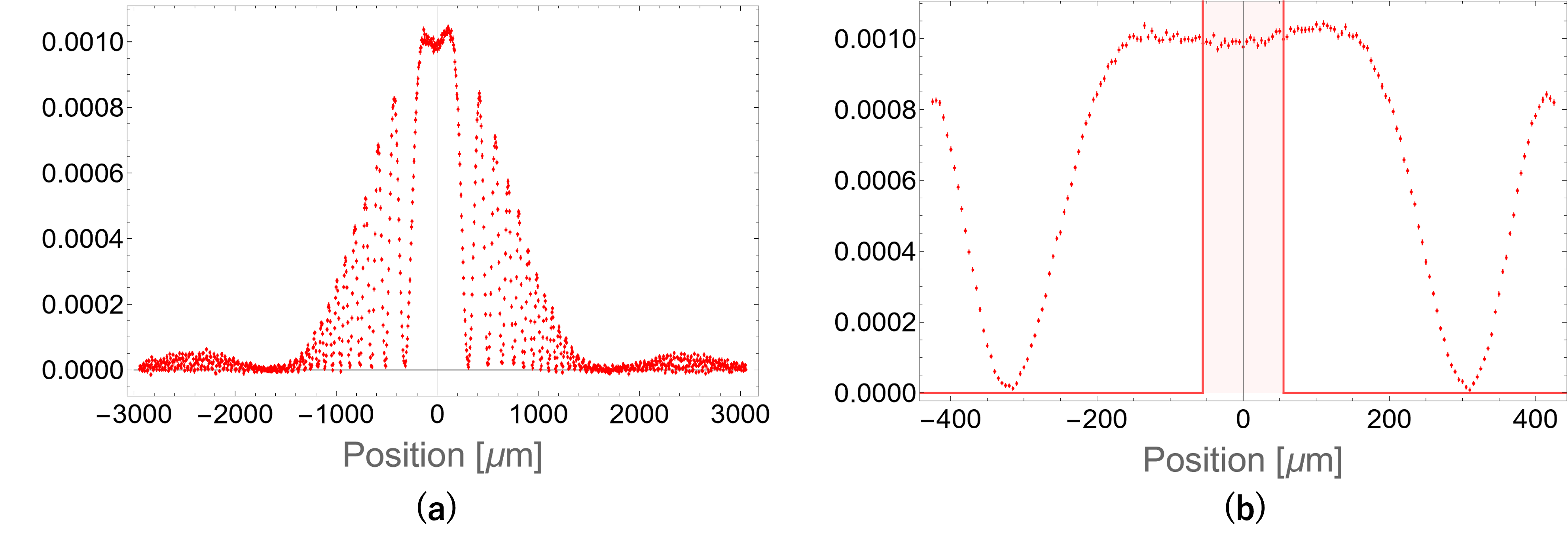}
\captionsetup{justification=raggedright,singlelinecheck=false}
  \caption{
  Experimental result at $t=t_M$. The vertical axis shows probability density in unit of $\mathrm{\mu m^{-1}}$. (a) shows the position distribution over the full range of 6 mm. (b) shows the vicinity of the interval $M$ at the center. The shaded area indicates the interval of width of $M = L+(cB/p)t_M = 110\:\mathrm{\mu m}$. 
  }
\label{fig:state_tM}
\end{figure*}

We can now look at the data obtained for the position distribution at $t=t_M$, which was taken $100 \: \mathrm{mm} \pm 0.25\:\mathrm{mm}$ downstream from the $t=0$ plane. Fig. \ref{fig:state_tM} shows the probability density of the position distribution. The interval $M$ shown in Fig. \ref{fig:state_tM} has a width of $L+(cB/p)\:t_M = 110\:\mathrm{\mu m}$. The probability $P(M)$ can be evaluated by summing over all counts oberved within this interval and dividing the result by the sum over all counts. The experimental value determined in this manner is given by $P_{\mathrm{exp}}(M) = 0.1092 \:\pm\:0.0002$. Comparison with the experimental results of the position and momentum measurements shows a violation of the particle propagation inequality by $P_{\mathrm{defect}}=0.0602\:\pm\:0.0010$. 

The full distribution of intermediate positions is shown in Fig. \ref{fig:state_tM} (a). It is easy to see that the phase differences between the position component and the momentum component result in an interference pattern with a visibility close to one. Using the quasi-probabilities of Eq.(\ref{eq:exp_state}), we can express the theoretically expected result as 
\begin{equation}
\begin{split}
    \bra{x_M} \hat{\rho} \ket{x_M} = & w_{L} |\braket{x_M|L}|^2 + w_{B} |\braket{x_M|B}|^2 \\[0.2cm] 
    & + w_{\mathrm{inter}}\left|\frac{\braket{x_M|L}\!\braket{B|x_M}}{\braket{B|L}}\right| \cos( \phi(x_M)),
\end{split}
\end{equation}
where $\phi(x_M)$ is the phase difference between $\ket{L}$ and $\ket{B}$ at $x_M$. Eq. (\ref{eqn:wf_pos_mom}) shows that $|\braket{x_M|L}|=|\braket{x_M|B}|$. This means that we can identify an envelope function defined by the maxima of the interference pattern, 
\begin{equation}
    \mathcal{E}(x_M) = \left(w_{L}+w_{B} + \frac{w_{\mathrm{inter}}}{|\braket{B|L}|}\right) A^2(x_M) ,
\end{equation}
where $A^2(x_M)$ represents the absolute values of the squared amplitudes, $|\braket{x_M|L}|^2 = |\braket{x_M|B}|^2$. 
\begin{figure}[h]
    \centering
    \includegraphics[width=\linewidth]{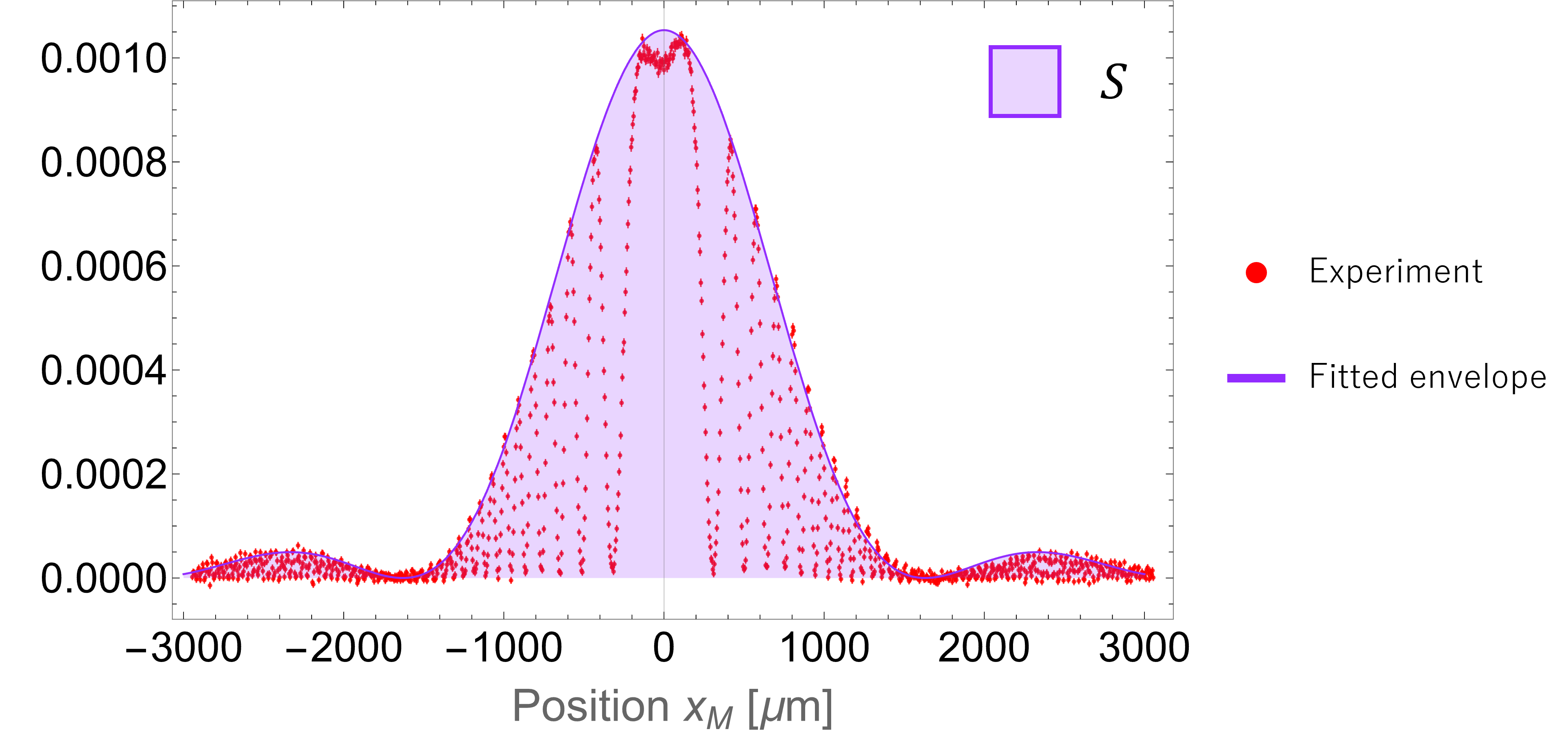}
    \captionsetup{justification=raggedright,singlelinecheck=false}
    \caption{
    Estimation of the area $S$ under the envelope. The envelope of the interference pattern is fitted using a squared sinc function and the area is determined from the height of the function and the location of the minima.
    }
    \label{fig:envelope}
\end{figure}
Fig. \ref{fig:state_tM} (a) indicates that the visibility of the interference fringes is close to one. In that case, the amplitude of the interference term is exactly equal to the contributions from $\ket{L}$ and $\ket{B}$, so that
\begin{equation}
\label{eq:vis1}
    \frac{w_{\mathrm{inter}}}{|\braket{B|L}|} \simeq w_{L} + w_{B}
\end{equation}
It is then possible to determine the quasi-probability $w_{\mathrm{inter}}=1-w_{L}-w_{B}$ from the area $S$ under the envelope $\mathcal{E}(x_M)$,
\begin{equation}
\begin{split}
    S = \int_{-\infty}^{\infty} &\mathcal{E}(x_M)\:dx_M\\[0.2cm] 
      & =2 (w_{L}+w_{B}).
\end{split}
\end{equation}
Fig. \ref{fig:envelope} shows the fit of the experimental data used to determine this area. The numerical result obtain from the data is
\begin{equation}
    S_{\mathrm{exp}} = 1.718 \pm 0.007.
\end{equation}
The estimated value of $w_{\mathrm{inter}}$ is $0.141$, close to the value obtained directly from $P_{\mathrm{exp}}(L)$ and $P_{\mathrm{exp}}(B)$ using $|\braket{B|L}|^2=0.0309$. The value of $|\braket{B|L}|$ obtained from Eq.(\ref{eq:vis1}) is approximately $0.164$, the square of which is $0.027$. This value is close to the one obtained for a slit width of $48 \:\mathrm{\mu m}$, providing additional evidence that this might be the actual slit width. 

We can now identify the contribution of interference to the value of $P(M)$. To do so, we first need to consider the contributions of $\ket{B}$ and $\ket{L}$. Due to the choice of $t_M$, these contributions are expected to be exactly twice as high as the contributions of $\ket{B}$ to $P(L)$ and of $\ket{L}$ to $P(B)$. Based on the fits of the diffraction patterns in the position and the momentum measurements, the contributions of $\ket{B}$ to $P(L)$ and of $\ket{L}$ to $P(B)$ are given by $0.0159$ and $0.0116$. The total contribution of $\ket{B}$ and $\ket{L}$ to $P(M)$ is equal to $0.0550$, leaving $P_{\mathrm{exp}}(M)-0.0550 = 0.0542$ to the effects of interference. As expected for constructive interference, the contributions are approximately equal. Using Eq.(\ref{eq:vis1}), the contributions to $P(M)$ can be approximated by
\begin{equation}
    P(M) \simeq 2 |\braket{B|L}|^2 (w_L+w_B) + 2 |\braket{B|L}| w_{\mathrm{inter}}.
\end{equation}
The contribution of interference is given by the last term of this equation. For $w_{\mathrm{inter}}=0.152$, a contribution of $0.0550$ is obtained with $|\braket{B|L}|^2=0.0327$ ($2|\braket{B|L}| = 0.3617$), consistent with the values obtained by fitting the diffraction patterns.

Even though the quasi-probability of interference is $100 \%$ localized in $L$ and in $B$, only a fraction of $2|\braket{B|L}|$ of it appears in $M$. This is why a quasi-probability of $w_{\mathrm{inter}}=0.152$ is sufficient to obtain a defect probability of $P_{\mathrm{defect}}=0.0602\:\pm\:0.0010$. The specific contribution of the interference effect can be found by subtracting the contribution of $0.0550$ to $P(M)$ from $w_{\mathrm{inter}}=0.152$. The result is $0.097$, where the difference of $0.037$ is accounted for by the excess probabilities of $\ket{L}$ and $\ket{B}$. Overall, about $64\%$ of $w_{\mathrm{inter}}$ contributes to $P_{\mathrm{defect}}$. 

\subsection{\label{sec:sublevel2} Negativity of the Wigner function}

$w_{\mathrm{inter}}$ is a quasi-probability because it represents interference patterns that can take on negative values. The Wigner function $W(x,p)$ is a more detailed quasi-probability, where each phase space point $(x,p)$ is associated with particle motion in a straight line. If a minimal joint probability of $P(L)+P(B)-1$ resulted in an equally high value of the quasi-probability $W_{LB}$, the particle propagation inequality could never be violated. However, Eq.(\ref{eq:WLBmax}) limits this value to $LB/(\pi \hbar)$. For the detection intervals of $55 \:\mathrm{\mu m}$, $W_{LB}\leq 0.074$. This is significantly lower than the minimal joint probability of the interference term, $w_{\mathrm{inter}}=0.152$. According to Eq.(\ref{eq:negative_wigner}), the integral over the Wigner function over the phase space region outside of both intervals $L$ and $B$ has an upper bound of 
\begin{equation}
    W_{\mathrm{out}}\leq -0.095.  
\end{equation}
Since the contributions of $\ket{L}$ and $\ket{B}$ do not contribute to $W_{\mathrm{out}}$, this negative quasi-probability originates entirely from the interference term with $w_{\mathrm{inter}}=0.152$. In the cross-shaped region of $W_{\mathrm{in}}$, the interference term must therefore contribute a positive quasi-probability of at least $0.247$ to the interference term of $w_{\mathrm{inter}}=0.152$. 

The high negative value of $W_{\mathrm{out}}$ is already more than sufficient to explain the violation of the particle propagation paradox. Indeed, the defect probability of $P_{\mathrm{defect}}=0.0602\:\pm\:0.0010$ is significantly lower than the minimal negativity required by the limitation of $W_{LB}$. According to Eq.(\ref{eq:defect}), the difference originates from a positive quasi-probability of
\begin{equation}
    W_{diag} \geq 0.035.
\end{equation}
This contribution must be positive whenever $P(M)>LB/(\pi \hbar)$. The role of the negativity of $W_{\mathrm{out}}$ is to allow for a quasi-probability of $W_{\mathrm{in}}>1$ that contributes more to the values of $P(L)$ and $P(B)$ than $W_{diag}$ contributes to $P(M)$. As we have seen in the analysis above, it is the constructive interference within the intervals $L$ and $B$ that violates the inequality. In $P(M)$, the interference contributes the same as $\ket{L}$ and $\ket{B}$. In the end, it appears as if photons that contribute to the interference effect show up in $L$ and in $B$, but somehow show up in $M$ with much lower probability.

\section{\label{sec:level5}Conclusions}

The experimental results we obtained using a Sagnac interferometer to prepare a superposition of position and momentum not only confirm the inequality violation that is the hallmark of the particle propagation paradox, but also reveal the statistics of interference between these two very different components. Our analysis makes use of the linearity of quantum statistics to divide the observed events into contributions characterized by positive quasi-probabilities. Here, the interference term appears as a separate contribution, added to the position state and the momentum state, each of which have their own quasi probability. 

The interpretational problem of this analysis is that we do not know what the relation between the three different measurements is. If photons were point particles moving along trajectories, each individual photon would appear at exactly one point in each distribution, and these three points would have to be connected by the trajectory of the photon. Essentially, this would be a hidden variable theory as described by Bohmian mechanics \cite{bohm}. However, Bohmian mechanics is at odds with the linearity of quantum mechanics. As pointed out by von Neumann, any hidden variable trajectory would have to place the same particle in different contributions to the density matrix at different points along the trajectory \cite{Golub}. Effectively, the particle would have to jump between orthogonal subspaces of Hilbert space to achieve a determinism completely different from the one described by the evolution of the wavefunction. 

Here, we take the approach suggested by von Neumann and assume that the photons detected in the diffraction pattern of position necessarily belong to the sub-ensemble described by $\ket{B}$, just as the photons detected in the diffraction pattern of momentum necessarily belong to the sub-ensemble described by $\ket{L}$. We must then conclude that the remaining photons belong to the interference pattern. However, this interference pattern takes negative values in the density observed at the intermediate position, making a realist interpretation impossible. How then should we think about the way in which particles travel through space? It has been suggested that particle propagation is non-local, indicating that the problem is closely related to entanglement and the violation of Bell's inequalities \cite{Aguilar}. The question is whether experimental data can explain the nature of this non-locality \cite{Rafsanjani}. Our analysis identifies three distinct sub-ensembles and demonstrates that the deviation from motion in a straight line originates from negativities in the quasi-probabilities associated with the interference term between position and momentum. We therefore conclude that the problem lies with the role of the measurement. When the position at $t=0$ is measured, it is impossible to identify the position at $t=t_M$ or the momentum of the particle, not because of a technical problem, but because such a position or momentum does not exist. Instead, we should consider different causal connections. In particular, wave-like propagation suggests that particles might be delocalized when they are not ``detected'' at a specific location. We have recently obtained experimental evidence that this might indeed be the case \cite{Fukuda}.

In summary, we have applied three different measurements to a superposition of position and momentum and shown that we can separate the contributions of the position state $\ket{L}$, the momentum state $\ket{B}$, and the interference of the two. The relation between the three measurements indicates that the photons detected in the three different measurements might not be the same photons. The only way to identify point particle trajectories is to replace the linearity of the density matrix with unobservable nonlinearities. The experimental evidence suggest that there is no joint reality of measurement outcomes connecting the measurements performed at different points along the axis of propagation.

\section*{Acknowledgements}
This work was supported by ERATO, Japan Science and Technology Agency (JPMJER2402).

\section*{Data availability}
The data that support the findings of this study are available in Zenodo at \url{https://doi.org/10.5281/zenodo.19199833}

\nocite{*}

\bibliography{apssamp}

\end{document}